\documentclass[useAMS,usenatbib,usegraphicx]{mn2e}

\title[CLASS: source selection and observations]
{The Cosmic Lens All-Sky Survey:
I. Source selection and observations}

\author[S.T. Myers et al.]{S.T. Myers$^{1,2}$,
N.J.~Jackson$^{3}$,
I.W.A.~Browne$^{3}$,
A.G.~de Bruyn$^{4,5}$,
T.J.~Pearson$^{6}$,\newauthor
A.C.S.~Readhead$^{6}$,
P.N.~Wilkinson$^{3}$,
A.D.~Biggs$^{3}$,
R.D.~Blandford$^{6}$,
C.D.~Fassnacht$^{7,1,6}$,\newauthor
L.V.E.~Koopmans$^{6,3,4}$,
D.R.~Marlow$^{2,3}$,
J.P.~McKean$^{3}$,
M.A.~Norbury$^{3}$,
P.M.~Phillips$^{3}$,\newauthor
D.~Rusin$^{2,8}$,
M.C.~Shepherd$^{6}$
and C.M.~Sykes$^{3}$ \\
$^1$National Radio Astronomy Observatory, P.O. Box O, Socorro, 
   NM 87801, USA \\
$^2$Department of Physics and Astronomy, University of Pennsylvania,
   209 S. 33rd St., Philadelphia, PA  19104-6396, USA \\
$^3$University of Manchester, Jodrell Bank Observatory, Macclesfield, 
   Cheshire SK11 9DL, UK \\
$^4$Kapteyn Astronomical Institute, Postbus 800, 9700 AA Groningen,
   The Netherlands\\ 
$^5$NFRA, Postbus 2, 7990 AA Dwingeloo, The Netherlands \\
$^6$California Institute of Technology, 105-24, Pasadena, CA 91125, USA \\
$^7$Space Telescope Science Institute, Baltimore, MD 21218, USA \\
$^8$Harvard-Smithsonian Center for Astrophysics, 60 Garden St., MS-51, 
   Cambridge, MA 02138, USA
}

\begin{document}

\maketitle

\begin{abstract}
The Cosmic Lens All-Sky Survey (CLASS) is an international
collaborative program which has obtained high-resolution
radio images of over 10000 flat-spectrum radio sources in
order to create the largest and best studied statistical sample of radio-loud
gravitationally lensed systems.  With this survey, combined with detailed
studies of the lenses found therein, constraints can be placed on the
expansion rate, matter density, and dark energy (e.g. cosmological constant,
quintessence) content of the Universe that are complementary to and
independent of those obtained through other methods.  CLASS is aimed at
identifying lenses where multiple 
images are formed from compact flat-spectrum radio sources, which should be 
easily identifiable in the radio maps.  Because CLASS is
radio-based,  dust obscuration in lensing galaxies is not a factor, 
and the relative insensitivity of the instrument to environmental conditions
(e.g. weather, ``seeing'') leads to nearly uniform sensitivity and resolution
over the entire survey.  In four observing ``seasons'' from
1994--1999, CLASS has observed 13783 radio sources with the VLA at 8.4~GHz
in its largest ``A'' configuration ($0\farcs2$ resolution).   When
combined with the JVAS survey, the CLASS sample contains over 16,000
images.  A complete sample of 11685 sources was observed, selected to have
a flux density of at least 30~mJy in the GB6 catalogue at 4.85~GHz (spanning
the declination range $0^\circ \leq \delta \leq 75^\circ$ and
$|b| \geq 10^{\circ}$, excluding the galactic plane) and 
a spectral index $\alpha \geq -0.5$ between the NVSS at 1.4~GHz and the
GB6.  A typical 30-second CLASS snapshot reached an rms noise level of
$0.4$~mJy. 
So far, CLASS has found 16 new gravitational lens systems, and the JVAS/CLASS
survey contains a total of 22 lenses. The follow-up of a small number of
candidates using the VLA, MERLIN, the VLBA, and 
optical telescopes is still underway.  In this paper, we
present a summary of the CLASS observations, the JVAS/CLASS sample,
and statistics on sub-samples of the survey. A companion paper presents
the lens candidate selection and in a third paper the implications for
cosmology are discussed. The source catalogues from the JVAS/CLASS
project described in this paper are available from
{\tt http://www.jb.man.ac.uk/research/gravlens/} .

\end{abstract}



\section{Introduction} \label{intro}

The Cosmic Lens All-Sky Survey (CLASS) is an international program to map more
than 10,000 radio sources in order to create the largest and best studied
statistical sample of radio-loud gravitationally lensed systems.  CLASS is
aimed at identifying lenses where multiple images are formed from compact
flat-spectrum radio sources.  The resulting lens configurations should be
easily identifiable in the radio maps.  CLASS is most efficient at finding
galaxy-mass lenses with arcsecond image separations, as these mass
distributions dominate the lensing optical depth.  In principle, CLASS can
detect more extreme lensed systems with larger separations due to clusters of
galaxies (Phillips et al.\ 2001), 
as well as smaller-separation (160-300mas) lenses (Augusto, Wilkinson \&
Browne 1998; Augusto \& Wilkinson 2001).
Setting limits on gravitational lensing on these smaller and larger angular
scales is a secondary goal of the survey.

\begin{table*}
\begin{tabular}{llccl}
\small
Lens & Session & Max. Sep. & \# images & References \\
B0128+437 & CLASS-1 & $0\farcs54$ & 4 & Phillips et al.\ 2000a \\
{\bf B0218+357} & JVAS    & $0\farcs33$ & 2 + ring & Patnaik, Porcas
\& Browne 1995; Biggs et al.\ 1999 \\
MG0414+054 & MG, JVAS & $2\farcs09$ & 4 & Hewitt et al.\ 1992 \\
{\bf B0445+123} & CLASS-2 & $1\farcs33$ & 2 &  Jackson et al.\ 2002, in
prepartation \\
{\bf B0631+519} & CLASS-1 & $1\farcs16$ & 2 & York et al.\ 2002, in
prepartation \\ 
{\bf B0712+472} & CLASS-1 & $1\farcs27$ & 4 & Jackson et al.\ 1998 \\
B0739+366 & CLASS-2 & $0\farcs54$ & 2 & Marlow et al.\ 2001 \\ 
{\bf B0850+054} & CLASS-3 & $0\farcs68$ & 2 & Rusin et al.\ 2002b, in
preparation \\
B1030+074 & JVAS    & $1\farcs56$ & 2 & Xanthopoulos et al.\ 1998 \\
B1127+385 & CLASS-2 & $0\farcs70$ & 2 & Koopmans et al.\ 1998 \\ 
{\bf B1152+199} & CLASS-3 & $1\farcs56$ & 2 & Myers et al.\ 1999; Rusin et al.\ 2002a \\
{\bf B1359+154} & CLASS-3 & $1\farcs65$ & 6 & Myers et al.\ 1999; Rusin et al.\ 2001b \\
{\bf B1422+231} & JVAS    & $1\farcs28$ & 4 & Patnaik et al.\ 1992b \\
B1555+375 & CLASS-2 & $0\farcs43$ & 4 & Marlow et al.\ 1999 \\ 
B1600+434 & CLASS-1 & $1\farcs39$ & 2 & Jackson et al.\ 1995 \\ 
{\bf B1608+656} & CLASS-1 & $2\farcs08$ & 4 & Myers et al.\ 1995; Snellen et al.\ 1995 \\ 
{\bf B1933+503} & CLASS-1 & $1\farcs17$ & 4 + 4 + 2 & Sykes et al.\ 1998 \\
B1938+666 & JVAS    & $0\farcs93$ & 4 + ring & King et al.\ 1997 \\
{\bf B2045+265} & CLASS-2 & $1\farcs86$ & 4 & Fassnacht et al.\ 1999a \\
B2108+213 & CLASS-4 & $4\farcs60$ & 2 & McKean et al.\ 2002, in preparation \\
{\bf B2114+022} & JVAS    & $2\farcs57$ & 4 & Augusto et al.\ 2001; Chae, Mao
\& Augusto 2001 \\
{\bf B2319+051} & CLASS-2 & $1\farcs36$ & 2 & Rusin et al.\ 2001a \\
\end{tabular}
\caption{ JVAS/CLASS Lenses \label{tbl:lens}
The separation for lenses with more than 2 components is
the maximum separation.  The number of images listed is the number of
radio components believed to be lensed images,
not counting sub-components seen on VLBI scales.  The principal references are
given for each lens. The thirteen lenses from the CLASS Complete Sample
(section 2.2) are shown in boldface.
}
\end{table*}

The Very Large Array\footnote{The National Radio Astronomy Observatory is
operated by Associated Universities, Inc., under cooperative agreement with
the National Science Foundation.} (VLA) was used as the primary
instrument for the CLASS survey.  In its largest ``A'' configuration, the VLA
provides high-quality snapshot images with angular resolution of 0.2--0.3 
arcseconds at an average observing frequency of 8.46~GHz.  In the four primary
phases of CLASS carried out from 1994 -- 1999, a total of 13832 target radio
sources were mapped. From among these, hundreds of multiple-component sources
were identified and followed up with higher resolution using MERLIN (50~mas
resolution) and the VLBA (2~mas resolution), and with multi-wavelength 
observations using a number of instruments.  
So far, CLASS has discovered 16 new radio-loud gravitational lens systems,
which when added to the 6 lenses from JVAS makes a total of 22
confirmed lens systems in JVAS/CLASS (Table~\ref{tbl:lens}).
Some follow up observations are still
underway but it is doubtful that more multiple image
systems will be found in the sample.
For example, we have yet to conclusively identify the CLASS-2 source CLASS
B0827+525, a $2\farcs82$ double, as either a `dark lens' or binary radio-loud
quasar (Koopmans et al.\ 2000b).  The details of this follow-up
program and the verified lenses will be given in Paper~2.

There have been previous VLA-based lens surveys: the MIT-Green Bank (MG)
survey (Burke 1989), and the Jodrell Bank VLA astrometric source survey (JVAS)
(Patnaik et al.\ 1992a; Browne et al.\ 1998; Wilkinson et al.\ 1998).  
JVAS sources, like CLASS sources, were required to
have flat radio spectra, but the JVAS survey was restricted to 5-GHz radio
flux densities $\geq 200\,$mJy, compared to the 25--30$\,$mJy limit of CLASS. 
JVAS is thus
effectively a bright sub-sample of CLASS. The main difference between
JVAS/CLASS and the MG survey is that the latter has no spectral index
selection, and so also picks up sources dominated by steep-spectrum lobe
emission. Of the $\sim 5000$ target sources observed in the two previous
surveys, five multiply imaged lens systems were found in the MG survey and
six in JVAS, of which one was a rediscovery of one of the MG lenses.  CLASS
has more than doubled the number of radio-discovered lens systems known, and
JVAS/CLASS as a whole constitutes the most comprehensive single sample of
lenses for use in astrophysical applications.  In addition to these Northern 
celestial hemisphere surveys, the VLA is also being used to carry out a 
gravitational lens survey in the Southern sky (Winn et al.\ 2000).

One of the most important uses of galaxy-mass lenses is in the
determination of cosmological parameters.  In particular, time delays
measured between the components in multiple image systems with reliable
mass models can directly determine the angular-diameter distances to the lens
and lensed 
objects, and therefore the Hubble constant $H_0$.  Because the lensing
equations depend on a ratio of distances, the delays are not as sensitive to
the matter density $\Omega_{\rm m}$ and cosmological constant
$\Lambda_0$.  Of course, the success of the method depends upon the
presence of measurable variability in the lensed source, and the
ability to construct a well-constrained mass model for the lens, which
in turn should ideally consist of a single deflector well-fit by a
spherical or ellipsoidal potential.  Compact flat-spectrum lensed radio
sources are more likely to be variable and are easily recognized as
multiply-imaged. Steep spectrum, extended lensed sources can provide
more constraints on the lensing potential, but are unlikely to be variable
and are hard to distinguish from complex unlensed radio lobes.
Thus, CLASS is aimed at filtering out the latter in preference for the former.
Unfortunately, only a few lenses found in CLASS will satisfy these
criteria and be suitable as cosmological standards.  
The CLASS group has successfully monitored the quadruple lens CLASS B1608+656
(Fassnacht et al.\ 1999b; Koopmans \& Fassnacht 1999, Fassnacht et al.\ 2002),
and all three independent time delays between the four components have been
measured.  For the current standard cosmology ($\Omega_{\rm m}=0.3$,
$\Lambda_0=0.7$) a Hubble constant $H_0=$ 61--65 (galaxy centroid positions)
$\pm 2$ ($2\sigma$ statistical) $\pm 15$ (systematic) km$\,$s$^{-1}$Mpc$^{-1}$
has been derived (Fassnacht et al.\ 2002).  
In addition, the double image systems
CLASS B1600+434 (Koopmans et al.\ 2000a, Koopmans \& de Bruyn 2000)
and JVAS B0218+357 (Biggs et al.\ 1999) were monitored and time delays
measured, giving Hubble constant values of 57$^{+14}_{-11}$ and
69$^{+13}_{-11}$ km$\,$s$^{-1}$Mpc$^{-1}$ respectively (errors are $2\sigma$
statistical errors assuming singular isothermal ellipsoid mass models).
The mass profile modeling for these systems is being further refined, and 
new results will be presented in upcoming papers.

The observed lensing rate, and to a lesser extent the distribution of
image separations and redshifts of the lenses, provides constraints upon the
cosmology, particularly the differential volume of the universe versus
redshift which is largely controlled by the matter density $\Omega_{\rm m}$
and the dark energy density $\Lambda_{0}$ and its equation of state
parameter $w$ ranging between $w=-1$ for a cosmological constant and $w=0$ for
non-relativistic matter.  A well-defined sample selection
is necessary to understand the statistics of the lensing, which is in turn
necessary to constrain $\Lambda_0$ (Turner, Ostriker \& Gott 1984; Kochanek
1996).  For this purpose it is vital that all the
lenses in the observed sample must be identified.  Because CLASS is
radio-based, complications arising from dust obscuration in lensing galaxies
do not occur; and because it targets flat-spectrum radio sources with
intrinsically simple radio structures, real lenses are easy to identify. 
Furthermore, the external conditions do not strongly affect the observations
and the survey was carried out with nearly uniform sensitivity and resolution.
All these factors make it the best sample overall for statistical studies of
lenses.  However, interpretation of the statistical results from any
observational survey in the context of theoretical models requires the control
of selection effects in the data sample.  In the case of CLASS, this means
studying the effects of the lensing on the parent sample of the survey, the
resulting image geometries, the ability of the automatic mapping 
procedure to identify the
lensed images, and possible loss of lenses from the survey.  In addition,
the CLASS sources suffer from the drawbacks generic to radio-selected samples
of having poorly constrained redshift distributions and source luminosity
functions, which must be dealt with both statistically and through follow-up
optical observations.

Data from CLASS also provide important astrophysical constraints upon the
mass distribution within the lensing galaxies themselves (e.g.\ Kochanek
1995).  Koopmans \& de Bruyn (2000) observed the signatures of microlensing
in the radio light curves of the two images in the system CLASS B1600+434,
and deduce that a significant fraction of the mass in the inner
dark-matter halo of the lensing galaxy is in the form of compact objects.
Rusin \& Ma (2001) use the absence of faint "odd" images in deep radio maps of
CLASS lenses to place a lower limit on the inner mass profiles of lensing
galaxies.  Rusin \& Tegmark (2001)  
consider the frequency of quadruple image systems in
CLASS and attempt to derive a consistent set of astrophysical mechanisms that
explain this statistic as well as other observational constraints.  
Keeton \& Madau (2001) and Phillips et al.\ (2001) constrain the concentration of cluster-mass dark
matter halos based on the paucity of wide-separation lens systems in the CLASS
sample.  Cohn et al.\ (2001) use the lensing model for CLASS
B1933+503 to constrain density profiles.  Rusin et al.\ (2002a) use
high-resolution radio and optical observations of CLASS B1152+199 to model
the galaxy mass profile.  Finally, observations of the six-image system CLASS
B1359+154 was used to place constraints on the mass distribution within the
small group of galaxies responsible for the lensing (Rusin et al.\ 2001b).  
Astrophysical applications of lensing such as these
studies are as important as the cosmological uses outlined above, and may
ultimately provide critical clues to the nature and properties of dark matter
in galaxies and groups.

The success of the CLASS survey, and the unprecedented observing and analysis
speed, was made possible through automation of the observation scheduling and
mapping analysis pipeline.  Specialized auto-mapping software (see
\S~\ref{obs}) was developed to carry out these tasks.  This has allowed us to
observe an average of one target source per minute with the VLA, and then
later to map the calibrated data at a similar (now much faster) rate.

In addition to the lens candidates and lensing-related applications, the CLASS
survey provides a solid body of radio information for all the objects.  
Astrometric (35 mas) positions for the sample are determined, which extends the
MERLIN/VLBA phase-reference source network in the northern sky to a spacing of
one to a few degrees (except in the galactic plane).  CLASS has also proven to
be useful in studies of active galactic nuclei.  For example, JVAS and CLASS
have been used to identify samples of high-redshift quasars (Snellen et al.\
2001, 2002).  CLASS observations have supplemented lower frequency surveys in
identifying young radio sources (Snellen et al.\ 2000).  March\~{a} et al.\
(2001) used CLASS to identify a large sample of low-luminosity blazars.

The JVAS/CLASS survey is now complete. This is the first in a series of papers
which describe the survey and its results, and covers the initial
observations, data reduction and properties of the survey.  A second
companion paper (Browne et al.\ 2002) will describe the selection and
followup of lens candidates. A third paper (Chae et al.\ 2002) presents
the cosmological results, in particular the derived constraints on the 
matter density and the dark energy density and its equation of state.

\section{The survey sample} \label{sample}

To achieve our goal of obtaining $\sim 20$ new simple lenses suitable for
cosmological measurements, we must start with a parent sample of more than
10000 flat-spectrum sources.  Moreover, to be easily followed up with
higher-resolution radio interferometers such as Multi-Element
Radio-Linked Interferometer Network (MERLIN), the NRAO Very Long Baseline
Array (VLBA) and the European VLBI network (EVN), the sources should have
total compact emission of at least 20~mJy. Note that in a lensed system this
flux density can be split into four or more images.

To obtain compact multiply-imaged components, we target sources with
relatively flat radio spectra.  We use the convention
$S \propto \nu^\alpha$,
where $\nu$ is the frequency of observation.  We classify as ``flat-spectrum''
those sources with a spectral index of $\alpha \ge -0.5$ between
4.85~GHz and a lower frequency.  Technically, we start with our 4.85~GHz
sample and reject ``steep-spectrum'' sources that have $\alpha < -0.5$ versus
a lower frequency, as in some stages of the survey targets were kept in the
sample when low frequency measurements were unavailable.

The CLASS observations were taken over five years: CLASS-1 in 1994, CLASS-2 in
1995, CLASS-3 in 1998 and CLASS-4 in 1999.  Table~\ref{tbl:stat} shows the
progression of the survey from JVAS through the end of CLASS.  The
selection for JVAS is described by Patnaik et al.\ (1992a). The CLASS
survey selection is complicated, and we describe it here in some detail
as it affects the use to which the data products may be put. 
We also define and describe a simpler subset (the CLASS complete sample).

Our starting point for source selection in CLASS-1 and CLASS-2 was the 87GB
version of the Green Bank Survey at 4.85~GHz (Gregory \& Condon 1991).
However, the updated survey GB6 (Gregory et al.\ 1996) became available and
was adopted as the parent sample; in this survey the source positions and flux
densities are better determined than in 87GB.  Furthermore, in the later
stages of the CLASS survey, we adopted the NRAO-VLA Sky Survey (NVSS, Condon
et al.\ 1998) as the low frequency survey to define the spectral selection.
We now describe the observation sessions and the respective selection criteria
in detail.

\begin{table*}
\begin{tabular}{ccccccccc}
Session & Dates & $N_{\rm obs}$ & $N_{\rm det}$ & \% detect & 
$N_{\rm mult}$ & \% mult & BW MHz & Selection \\
   JVAS     & Feb 1990 -- Dec 1992 & 2720 & 2613 & 96.1\% 
            & 456 & 17.5\% & 25/50 & GB \\
   CLASS-1a & Feb 1994 -- May 1994 & 2550 & 2199 & 86.2\% 
	    & 644 & 29.3\% & 25 & Texas \\
   CLASS-1b & ''                   & 670 & 624 & 93.1\% 
            & 65 & 10.4\% & 50 & WENSS \\
   CLASS-2a & Aug 1995 -- Sep 1995 & 2337 & 2150 & 92.0\% 
            & 176 & 8.2\% & 50 & WENSS \\
   CLASS-2b & ''                   & 2098 & 1813 & 86.4\% 
            & 230 & 12.7\% & 25 & Texas/WENSS \\
   CLASS-3a & Feb 1998 -- May 1998 & 2338 & 2043 & 87.4\% 
            & 230 & 11.3\%  & 50 & NVSS \\
   CLASS-3b & ''                   & 2058 & 1823 & 88.6\% 
            & 249 & 13.7\% & 50 & NVSS \\
   CLASS-3c & ''                   & 657 & 575 & 87.5\% 
            & 125 & 21.7\% & 50 & NVSS \\
   CLASS-4  & Aug 1999             & 1075 & 753 & 70.0\% 
            & 371 & 49.3\% & 50 & NVSS \\[2ex]
\sc Total               & & 16503 & 14593 & 88.4\% & 2546 & 17.4\% & & \\
\sc Complete Sample  & & 11685 & 10906 & 93.3\% & 1558 & 14.3\% & & \\
\end{tabular}
\caption{JVAS/CLASS Statistics \label{tbl:stat}.
These numbers reflect pointings that differ by more than
$120\arcsec$. Duplicate pointings have been excluded. Note that a
multiple source will be recorded in cases where bad data or mapping
errors have resulted in a spurious secondary.  The ``\% mult'' column 
refers to the fraction of detected sources with multiple components of
any flux density including these spurious multiples --- if restricted to 
2~mJy or brighter the total is 15.1\% of sources detected for all
sources and 11.6\% in the complete sample.}
\end{table*}

\subsection{CLASS-1 and CLASS-2} \label{sample:class1and2}

In the first phase of CLASS, observed with the VLA in February--May 1994, 3220
target fields were mapped.  The sources were selected from the 87GB
catalogue (Gregory \& Condon 1991) which contains 54579 sources with 4.85~GHz 
flux densities of 25~mJy or greater in the declination range 
$0^\circ \leq \delta \leq 75^\circ$.  We imposed a galactic latitude limit of
$\vert b \vert > 10^\circ$.  
Spectral selection was based on two-point spectral indices
versus the 325 MHz Westerbork Northern Sky Survey (WENSS, Rengelink et al.\
1997), the 365~MHz Texas Survey (Douglas et al.\ 1980), and the 1.4~GHz Green
Bank survey (White \& Becker 1992).  When available we used the WENSS since
the flux densities are reliable and sources have a positional accuracy of
$\sim 5$ arcsecond.   At the time of CLASS-1, only a few ``mini-survey'' areas
were complete, but by the time of CLASS-2, WENSS was complete above 
$\delta>30^\circ$. In CLASS-1 and CLASS-2, sources outside the area covered by
WENSS at that time were selected using the Texas 365~MHz catalogue.

WENSS has a flux density limit of 15 -- 20 mJy while the 87GB limit is $\sim
25$~mJy. Target sources were selected with $\alpha \geq -0.5$ between 325~MHz
and 4.85~GHz.  There were 670 CLASS-1 sources selected using WENSS (designated
as CLASS-1b).  The remaining CLASS-1 sources were selected on the basis of
flux densities from the Texas 365~MHz catalogue.  Because the Texas survey has
a flux density limit of 250~mJy at 365~MHz, we relaxed the limits to 50~mJy or
greater at 4.85~GHz and $\alpha \geq -0.6$ between 325~MHz and 4.85~GHz.  All
sources meeting the criteria were observed for $\delta \geq 45^\circ
20^\prime$, and down to $\delta \geq 35^\circ$ south of the galactic plane.
In all, 2550 sources were observed with this selection (designated CLASS-1a).
Note that the Texas survey only covers the region $\delta < 71\fdg6$.  North
of this declination, there was effectively no spectral selection in CLASS-1a.

In CLASS-2, we were able to select more sources against WENSS as that survey
neared completion.  In all, 4435 more targets were observed.  The CLASS-2a
session observed 2337 targets that were nearly all selected using WENSS.  
The CLASS-2b observations mostly targeted sources below the WENSS southern 
limit ($+30^\circ$) and were selected using the Texas survey and the White \&
Becker 1.4~GHz catalogue (which has a flux density limit of 100~mJy at
1.4~GHz), 
excluding targets with $\alpha < -0.6$ between 365~MHz or 1.4~GHz and 4.85~GHz.

\subsection{CLASS-3, CLASS-4, and the CLASS Complete Sample} 
\label{sample:class3}

Our starting point for source selection in CLASS-1 and CLASS-2 had been the
87GB catalogue.  However, the improved Green Bank survey GB6 became available
in 1996.  This catalogue contains 75162 sources in the declination range
$0^\circ \leq \delta \leq 75^\circ$ with a 4.85~GHz flux density of
18~mJy or more and angular diameter $10\farcm5$ or smaller.
Since the source positions and flux densities are better determined
than in 87GB, GB6 was adopted as the primary finding list for CLASS-3, with
the intention of bringing the entire CLASS survey into line with this new
parent catalogue.  In addition, the NVSS became available, giving coverage of
the entire GB6 region at 1.4~GHz.  Use of NVSS created a sample with a uniform
rigorous spectral selection over virtually the whole northern sky.  This
GB6-targeted, NVSS spectrally-selected survey will be designated as the
``CLASS-NVSS Complete Sample'' or ``complete sample'', and naturally updates
the spectral selection used in JVAS as well as early CLASS.  We used the NVSS
version released as {\sc catalog39.fit} for sample selection.  This
release purported to be complete in the GB6 region ({\sc catalog40.fit}
is now available filling in holes in other regions of the sky).

The complete sample consists of 11685 targets with a GB6 flux density of
30~mJy or more, an NVSS source within $70\arcsec$ of the GB6 position, and a
spectral index $\alpha \geq -0.5$ versus NVSS. The spectral index is defined
by adding the flux densities of all NVSS sources within $70\arcsec$ of the GB6
position and ignoring all other NVSS sources.  The primary goals of CLASS-3
and CLASS-4 were to complete the sample.  A list of 4982 sources
for observation in CLASS-3 was drawn up from those sources not within
$2\arcmin$ of a CLASS-1 or CLASS-2 pointing center, JVAS target, or identified
with a known radio source (eg. from the VLA calibrator lists).

In addition, the CLASS-3 list was supplemented with a list of GB6 sources
(30~mJy or brighter), that have a WENSS source within $70\arcsec$ of the GB6
position, and spectral index $\alpha \geq -0.5$ versus WENSS.  This enables
the construction of a ``CLASS-WENSS Complete Sample'' and will catch
targets that have GHz-peaked spectra where the 1.4~GHz flux density is above
the spectral limit for inclusion in the CLASS-NVSS sample.

In total, 5053 new sources were observed in CLASS3 from Feb -- May 1998.  The
remaining sources that make up the complete sample were observed in CLASS-4,
with a total of 1075 targets observed in August 1999.  Note that most of the
CLASS-4 targets were near the lower flux density limit of the sample, entered
into the sample on the changeover to the NVSS, or were not detected in NVSS,
and thus there are a higher percentage of non-detections and extended or
multiple component sources (see \S~\ref{stats} and Table~\ref{tbl:stat}) in
this part of the survey.  The final combined CLASS and JVAS archive contains
16503 distinct pointings (more than $120\arcsec$ apart), of which 11685 form
the complete sample. The final complete sample covers an area of 4.96sr.

\begin{figure}
\includegraphics[width=8cm]{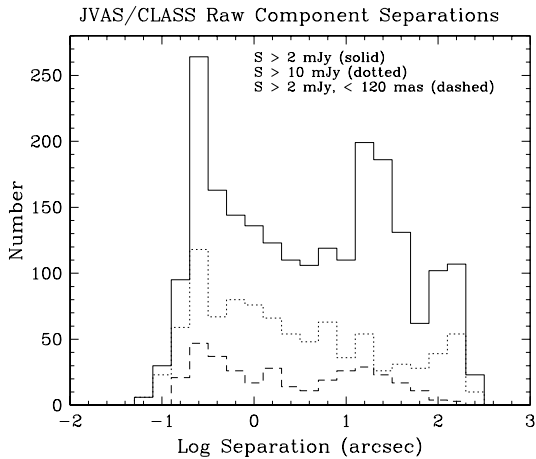}
\caption{Histogram of maximum component separations for the 2219 CLASS 
sources that have at least 2 components with individual flux densities of
2~mJy or more (solid).  Also shown are cuts at 10~mJy on component flux
densities for 938 sources (dotted), and with component sizes of 120~mas 
or smaller for 333 sources (dashed).  
\label{fig:dist}}
\end{figure}  

\section{CLASS VLA observations and analysis} \label{obs}

Two independent IFs of 25~MHz bandwidth (or 50~MHz with WENSS positions) were
centered at $8.415$ GHz and $8.465$ GHz (average $8.44$ GHz) for CLASS-1 and
CLASS-2.  In CLASS-3, which was observed in its entirety using NVSS positions
and 50~MHz bandwidth, the IF band centers were moved to $8.4351$~GHz and
$8.4851$~GHz (average $8.46$ GHz) in line with NRAO recommendations to avoid
radio frequency interference --- we will use $8.46$~GHz as the fiducial CLASS
frequency.  For all CLASS observations, an on-source dwell
time of 30 seconds was used, with $3.3$-second integrations.  A compact source
from the JVAS calibrator list was observed after every $n$-th target source
for phase calibration, where $n$ was chosen based on the weather conditions
prevalent during the session (for CLASS-1: $n=14$; CLASS-2: $n=8$; CLASS-3/4:
$n=10$--12).  The JVAS calibration sources have rms position errors in each
coordinate of 12~mas, so the resulting CLASS source positions should have
accuracies of about 20~mas (in stable observing conditions) and hence should
themselves be useful as potential phase calibration sources.  We were able to
observe one target source per minute, including the overhead from observing
calibration sources and slewing between sources.

\begin{figure}
\includegraphics[width=8cm]{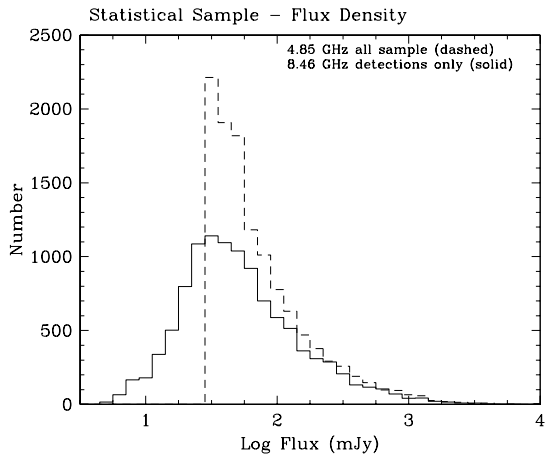}
\caption{Histogram of CLASS 8.46~GHz flux densities for detections 
in the complete sample with components merged (solid) and GB6 4.85~GHz
flux densities for the complete sample (dashed).  The GB6 flux
density limit of 30~mJy is easily seen in the dashed histogram.
\label{fig:fluxes}}
\end{figure}  

The gain calibration using the JVAS astrometric sources was carried out in the
Astronomical Image Processing Software
(AIPS) package, distributed by NRAO\footnote{http://www.cv.nrao.edu/aips}.  The
mapping, deconvolution and self-calibration were performed using the DIFMAP
package (Shepherd, Pearson \& Taylor 1994; Shepherd 1997).  
The imaging of the sources was completed in an automated fashion using a
DIFMAP script.  The average rate for processing with the script was one source
every two minutes.  This procedure yielded images of good quality for most of
the target fields. 

The relevant details of the calibration and mapping procedures adopted
for CLASS are discussed in the remainder of this section.

\subsection{Calibration and positional accuracy} \label{obs-pos}

Editing and calibration of the data was done using AIPS following
the standard procedure.  Initial processing was carried out at the
time of the observations in order to generate prompt candidate lists
for immediate follow-up where possible.  After the entire CLASS survey
was completed, final consistent editing/calibration was done for the full
dataset (including JVAS). The data were examined and edited using AIPS
TVFLG to remove obviously discrepant telescopes or baselines. The flux
density scale was set (AIPS SETJY) using the standard calibrators 3C48 and
3C286, for which the flux density scale at 8.41--8.48~GHz in the observations
was normalized to the Baars et al.\ (1977) values. Care was taken to ensure
that the flux densities on short baselines of these two sources approximated
well to the established values after the final correction was applied;
corrections were applied by hand if necessary.

The task AIPS CALIB was used to obtain telescope gain solutions on
calibrators, and the resulting solutions were examined. Bad solutions
were discarded by hand before interpolation by AIPS CLCAL into a final
gain solution. Calibration was applied to the CLASS sources and each
CLASS source was split from the file into an output disk UV-FITS file.

Fifteen sources were observed twice during the CLASS 1--3 sessions.
This provides a cross-check of the repeatability of our position
determinations.  These observations were calibrated independently, and did not
necessarily use the same phase calibrators.  For this set of sources, the rms
offset between the positions in the two epochs was found to be $35.2$~mas.
This is slightly higher than, but consistent with, our expected uncertainties
due to the JVAS position errors.

\begin{figure}
\includegraphics[width=8cm]{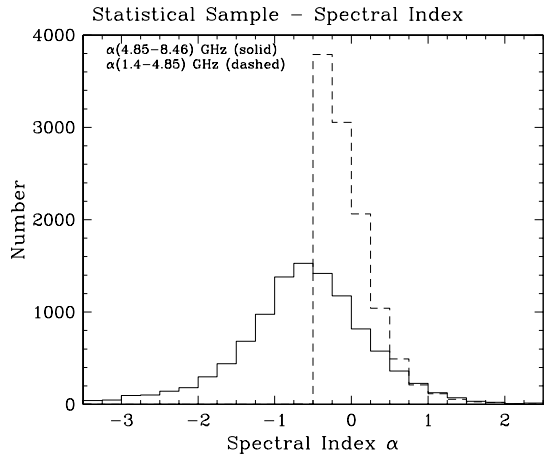}
\caption{Histogram of spectral indices between CLASS and GB6 (solid)
and between GB6 and NVSS (dashed).  Sources from the complete
sample detected in CLASS at 8.46~GHz are plotted (with any multiple components
merged).  The $\alpha \geq -0.5$ selection between 1.4 and 4.85~GHz is clearly
seen in the dashed histogram.
\label{fig:hista}}
\end{figure}  

\begin{table*}
\scriptsize
\begin{tabular}{lcccclcccrrr}
GB6J000007+081644&00000827&1.06&00 00 06.98&+08 16 46.2&980403&00 00 
07.0341&+08 16 45.040&31.0&18&0.76&23\\
GB6J000010+305556&00003093&1.28&00 00 10.10&+30 55 59.3&980403&00 00
10.0908&+30 55 59.420&26.4&87&0.31&50\\
GB6J000013+275142&00002786&1.31&00 00 14.76&+27 51 57.8&980403&00 00
14.8771&+27 51 57.577&29.9&21&0.00&14\\
GB6J000018+024812&00000280&1.05&00 00 18.60&+02 48 16.9&950829&00 00
19.2833&+02 48 14.657&85.2&36&0.00&85\\
GB6J000019+113918&00001165&0.95&00 00 19.50&+11 39 02.9&950902A\_FF1&00 00
19.5679&+11 39 20.718&29.2&47&0.00&-58\\
GB6J000029+471629&00004726&1.33&00 00 24.47&+47 16 09.8&980403&00 00
30.1085&+47 16 43.313&23.1&263&0.00&56\\
GB6J000026+030706&00000312&0.95&00 00 26.20&+03 07 21.0&950829&00 00
27.0230&+03 07 15.635&100.7&31&0.43&-51\\
GB6J000030+443127&00004451&1.35&00 00 26.97&+44 31 11.4&980403&00 00
26.8395&+44 31 11.700&7.9&90&0.73&-32\\
GB6J000035+291424&00012923&1.00&00 00 36.10&+29 14 12.0&950902A\_FF1&00
00 35.1294&+29 14 35.823&64.7&58&0.00&-28\\
GB6J000037+121357&00011223&NOSH&00 00 39.11&+12 13 53.7&980403&&&&&\\
GB6J000040+391758&00013929&1.05&00 00 41.20&+39 18 04.5&940404\_1&00 00
41.5259&+39 18 04.172&97.3&42&0.00&-38\\
&&&&&&00 00 41.4936&+39 18 05.092&21.4&122&0.74&6\\
GB6J000044+030744&00010312&1.03&00 00 45.30&+03 07 39.9&950829&00 00
44.3279&+03 07 54.199&63.3&45&0.39&-35\\
GB6J000048+121810&00011230&NOSH&00 00 49.41&+12 18 31.8&990816&&&&&\\
GB6J000049+325424&00013290&2.31&00 00 49.90&+32 54 24.0&990816&00 00
49.7361&+32 52 57.109&14.0&176&0.00&7\\
&&&&&&00 00 49.7358&+32 52 57.328&11.9&222&0.23&7\\
&&&&&&00 00 49.6133&+32 56 08.944&6.7&432&0.05&-6\\
&&&&&&00 00 49.5662&+32 56 08.820&5.5&1491&0.11&-87\\
GB6J000053+405401&2358+406&1.14&00 00 54.33&+40 53 56.1&900222&00 00
53.0817&+40 54 01.805&351.1&49&0.00&16\\
&&&&&&00 00 52.8036&+40 53 57.314&26.8&232&0.00&10\\
&&&&&&00 00 52.6903&+40 54 02.093&34.9&759&0.00&-48\\
GB6J000054+251605&00012527&1.12&00 00 56.04&+25 16 18.9&980314A&00 00
56.0910&+25 16 20.152&26.5&23&0.00&-14\\
GB6J000100+414932&00014182&1.77&00 01 01.40&+41 49 35.7&940305&00 01
01.4560&+41 49 27.929&23.9&163&0.41&30\\
&&&&&&00 01 01.5161&+41 49 31.101&11.4&185&0.00&-18\\
&&&&&&00 01 00.8887&+41 49 33.546&0.3&217&0.00&9\\
GB6J000107+242013&00012433&1.11&00 01 08.40&+24 20 06.9&950902\_FF1&00 01
07.8693&+24 20 11.799&51.0&36&0.00&-9\\
&&&&&&00 01 07.8555&+24 20 11.435&10.9&99&0.51&52\\
GB6J000109+191428&2358+189&1.08&00 01 08.86&+19 14 13.5&921017&00 01
08.6225&+19 14 33.818&504.2&59&0.71&32\\
GB6J000108+235307&00012387&NOSH&00 01 09.10&+23 52 39.9&950902\_FF1&&&&&\\
GB6J000107+024313&00010271&1.05&00 01 09.63&+02 43 10.0&980510&00 01
09.5365&+02 43 09.588&60.7&19&0.00&-47\\
GB6J000114+235801&00022396&1.21&00 01 14.85&+23 58 10.3&980314A&00 01
14.8643&+23 58 10.617&132.7&65&0.51&79\\
GB6J000115+061415&00020623&1.06&00 01 15.40&+06 14 12.0&950829&00 01
14.3441&+06 14 22.011&22.9&67&0.47&-59\\
GB6J000119+474202&00024769&1.47&00 01 19.60&+47 41 51.9&940305&00 01
19.0375&+47 42 00.717&100.5&50&0.13&28\\
&&&&&&00 01 19.5903&+47 42 00.525&3.8&113&0.00&-85\\
&&&&&&00 01 17.6877&+47 41 55.594&1.3&268&0.00&21\\
&&&&&&00 01 19.3247&+47 42 05.345&5.3&161&0.93&62\\
&&&&&&00 01 18.1426&+47 41 45.844&3.1&275&0.00&-9\\
&&&&&&00 01 18.3334&+47 42 01.115&2.7&134&0.00&-7\\
\end{tabular}
\caption{ \label{tbl:page1} First page of CLASS catalogue.
Column     1     (bytes 0-16):     GB6J name from the GB6 catalogue
(Gregory et al.\ 1996);
Column     2     (18-25):    Name in observing list; 
Column     3     (27-30):    Reliability number of map (see text). NOSH
indicates a non-detection; 
Column     4     (32-42):    Pointing position RA (J2000); 
Column     5     (46-55):    Pointing position Dec (J2000); 
Column     6     (60-70):    Observing epoch (90-92=JVAS, 94=CLASS1, 
                       95=CLASS2, 98=CLASS3, 99=CLASS4);
Column     7     (74-86):    Detected component position(s) RA (J2000); 
Column     8     (89-101):   Detected component position(s) Dec (J2000); 
Column     9     (103-111):  8.46-GHz flux density (mJy) of component(s); 
Column     10    (113-118):  Fitted major axis (mas) of component(s); 
Column     11    (120-123):  Fitted axis ratio of component(s); 
Column     12    (125-128):  Fitted position angle (degrees) of component.}
\end{table*}

\subsection{The auto-mapping procedure} \label{obs-check}

The {\it automap} algorithm can be outlined as follows. An initial
primary cycle is used to find peaks in a large image above some
signal-to-noise ratio (SNR) cutoff.  In a secondary cycle, small images 
are made around the location of each peak and a CLEAN window is created. 
In a tertiary cycle, deconvolution and self-calibration is carried out
iteratively.  The 
locations of the peaks (and thus the clean boxes) are stored for later use.

A series of parameters define the procedure. The dimensions (cellsize and
number of pixels) of the large primary image control the area around the
pointing position in which components can be identified. In practice bandwidth
smearing limits our effective field to about $1^\prime$ for the 50~MHz
observations which form the majority of the sample, and thus a $2048\times2048$
pixel search image was constructed with $0\farcs25$ cell size (giving a
$256^{\prime\prime}\times256^{\prime\prime}$ searchable inner quarter) and
super-uniform weighting.  This field 
was searched for pixels with values above a given SNR cutoff above which
sources are deemed to be reliably identified as real for further processing;
SNR=8 was adopted for CLASS.

Around each source, a secondary $2048\times2048$ image covering
$140^{\prime\prime}\times140^{\prime\prime}$ ($0\farcs0684$ pixels)
was then made using uniform weighting, of which the inner quarter
($70^{\prime\prime}\times70^{\prime\prime}$) was cleaned.  If a peak
is found in this secondary image above a second SNR cutoff (the ``field SNR
cutoff'', for which SNR=10 was adopted), a CLEAN box with a size 75\% of
the fitted CLEAN beam size is placed
around the image and 25 iterations of the CLEAN algorithm are carried out with
a gain of 0.05. The CLEAN component signal is subtracted from the data, and
the data inverted back into the $u$-$v$ plane, with the CLEAN procedure
repeated until a residual cutoff level (5\% of the peak or $6\sigma$) is
reached.  This means that we have a slightly higher sensitivity to secondary
features within this smaller field around the primary component than in the
larger search field, due to the extreme taper applied to the visibility data
input to the low-resolution search image (even accounting for the slightly
lower SNR cutoff in the small field), and thus are somewhat more likely to
find a faint source as a companion to a brighter source than on its own.

For each field identified and CLEANed, a phase-only self-calibration with
10-second solution interval is
performed and the CLEAN procedure is repeated. The CLEAN model is saved,
and a model-fitting procedure (DIFMAP MODELFIT) is run for 5 iterations
on the $u$-$v$ data.  The starting point for model-fitting is the clean boxes
found and stored during the iterative cleaning described above, with a 
Gaussian component assigned to each distinct CLEAN box (boxes must be
further than $0\farcs2$ apart).
The data products of the auto-mapping procedure outlined above are the
CLEAN model, the set of CLEAN windows, the list of MODELFIT components,
and the log-file generated by DIFMAP containing the commands executed plus
output information from the noise tests (this log-file can be executed
as a script in DIFMAP reproducing the original results).

We have adopted the Gaussian MODELFIT results as the best representation of
the DIFMAP analysis.  Each component has a best-fit position, major axis size,
elliptical Gaussian axis ratio and position angle.  In addition, uncertainties
for the fitted quantities are evaluated from the covariance matrix.  These
quantities can then be filtered to select candidates with multiple components,
and further filtered to find those with multiple compact components.  Note
that once distinct CLEAN boxes are laid down by the automapper, a component
is necessarily created that will be passed to the model-fitting.  The
model-fitting procedure may then move this component where it pleases,
possibly inside the exclusion box of $0\farcs2$, or possibly outward to
arbitrarily large separations.  There is no provision for deleting poorly
constrained components during model-fitting, and these must
be dealt with during the analysis phase.  For example, a moderate flux
density cutoff of 2~mJy per component, which corresponds to $5\sigma$ for
good data taken with 50~MHz bandwidth, is found to reject most of the
extraneous components.

The limitations of this algorithm include a restricted search field (only
$256\arcsec$ square around the pointing center in the inner quarter of
the large image), a relatively conservative
search SNR limit ($8\sigma$ to avoid many spurious detections) in this large
field, and dynamic range problems (often spurious multiple components are
found on sidelobes from bright sources).  However, the automatic mapping has
proved to be a reliable method of finding the best candidates quickly.

In some cases the automatic mapping program finds spurious structure or
sources. Without manual intervention it is difficult to find the cause
of many of these anomalies. However, the common symptom of such problems
is that the final peak flux density after cleaning and self-calibration is
much greater than the initial peak flux density. The ratio of these two
numbers, the ``reliability number'' (RN), should 
theoretically be around 1. In practice we have found that maps with
RN$>$1.5 usually indicate a data problem, and RN=2 almost certainly
indicates a spurious detection.

Because the goal of CLASS is to identify lens systems with multiple
compact components, it is crucial that we evaluate the effectiveness of
our automapper in finding such objects. To this end, we constructed fake
visibility data sets by adding pairs of point components (using the AIPS
task UVSUB) to actual CLASS fields in which no sources were detected by
the VLA. This was repeated over a grid of component separations 
($0\farcs15$--$10\arcsec$), total flux densities (30--100~mJy) and flux density
ratios (1:1--20:1) to simulate two-image gravitational lens systems. The
pairs were randomly oriented on the sky and placed up to 20 arcseconds
from the field center to mimic pointing inaccuracies. The data were then
passed through the automapper, and the resulting MODELFIT output was
analyzed. We find that for separations $\ga 300$~mas, the automapper
consistently picks up two compact components when the fainter component
has flux density $>2$~mJy. For a source with 30~mJy of total flux density,
this corresponds to a flux density ratio of approximately 15:1. As expected,
the identification of multi-component sources with larger total flux
density is complete to larger flux density ratios. The automapper becomes
increasingly inefficient at identifying multiple components as their
separation decreases below 300~mas. At 150~mas separation, almost all of
the data sets are fit to a single component, whose major axis increases
with decreasing flux density ratio.  

Finally, there were found to be minor irregularities in the
automapping for multiple component sources, with extraneous components being
generated due to the high sidelobes of the VLA snapshot beam, particularly
sidelobes from a source outside the small ($70^{\prime\prime}$) field.  
This effect can be seen in the histogram of maximum component
separations in the full sample is shown in Figure~\ref{fig:dist}.  The
secondary peak at around 20 arcseconds is almost certainly due to sidelobes,
as the dirty beam for a VLA snapshot has its main sidelobes, which have
an amplitude of 40\% of the peak, in a hexagonal pattern spaced at
$20^{\prime\prime}$ around the center.  
This was not deemed to be a serious problem for the
lens survey, as no true multiple sources would be lost, and simple 
filtering on the flux density and/or size of the modelfit components will
remove most of these cases, as shown by the dotted and dashed curves
in Figure~\ref{fig:dist}.  However, the reader must be warned that the
automatically generated catalogue still contains these spurious sources. 

\begin{figure*}
\vskip -1cm
\includegraphics[width=16cm]{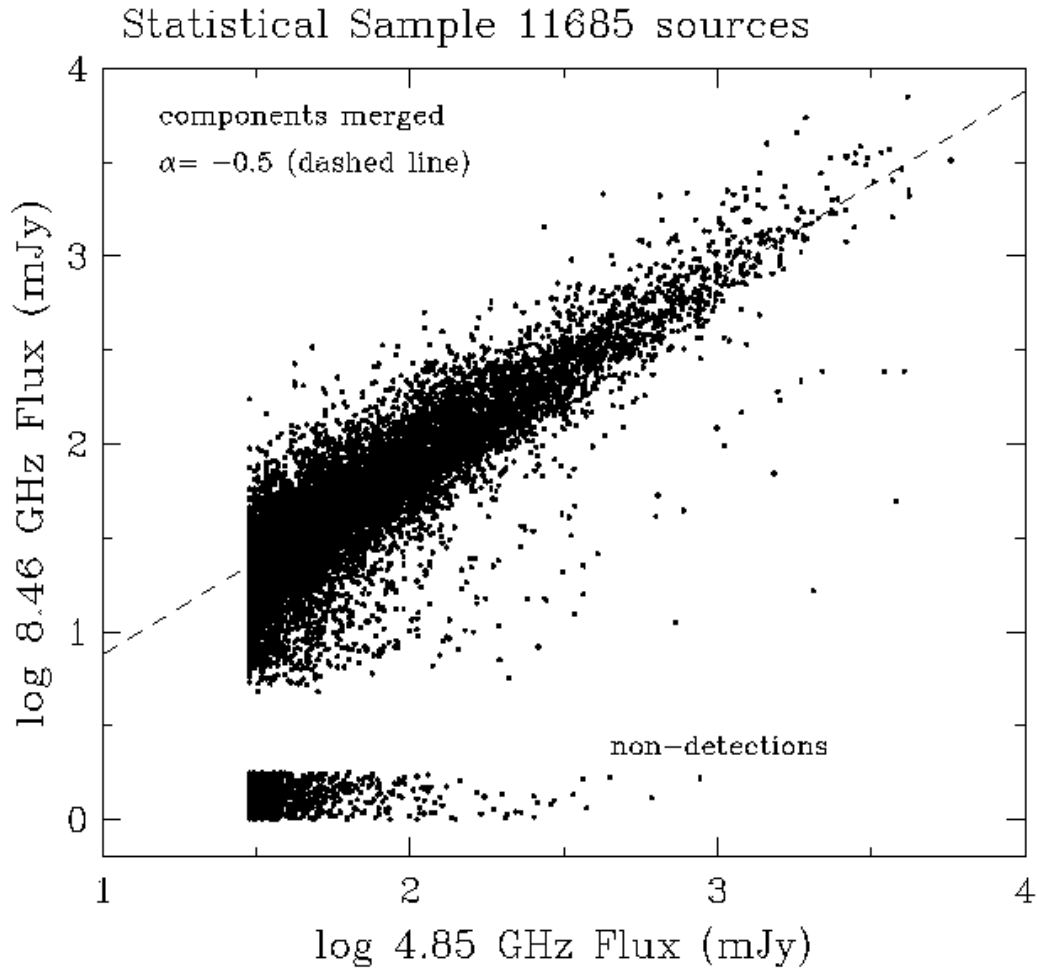}
\vskip -1cm
\caption{Observed JVAS/CLASS 8.46~GHz flux density versus GB6 4.85~GHz flux 
density, for the complete sample.  For sources found with
multiple components, the merged 8.46~GHz flux densities were used.  Sources
above the dashed line have apparent spectral indices between CLASS and GB6
flatter than $\alpha > -0.5$.  The bar of sources at the bottom represents GB6
targets where no CLASS source was detected (the assigned 8.46~GHz flux 
density is arbitrary and randomly spread to show the distribution with
4.85~GHz flux density).
\label{fig:jclvsgb6}}
\end{figure*}  

\begin{figure*}
\vskip -1cm
\includegraphics[width=16cm]{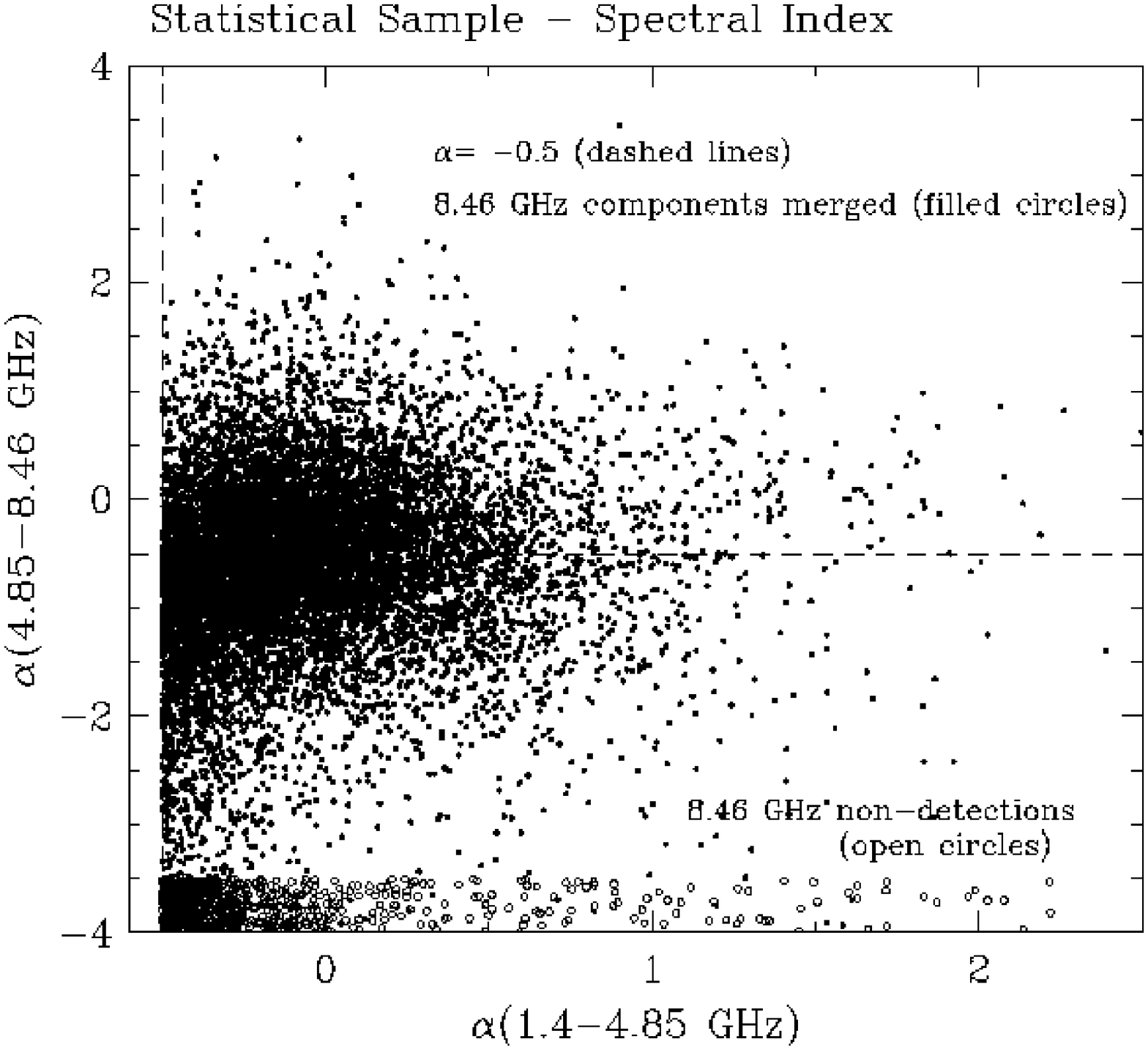}
\vskip -1cm
\caption{Spectral indices between CLASS and GB6 (8.46~GHz and 4.85~GHz) 
plotted versus GB6 to NVSS (4.85~GHz and 1.4~GHz) index 
for the complete sample. As before, multiple components are merged.  
The bar of open circles at the bottom represents sources in the 
sample for which no 8.46~GHz detection was made, plotted versus the selection
spectral index (y-values randomly spread to show the distribution).
\label{fig:alpha2}}
\end{figure*}  

\section{CLASS results and statistics} \label{stats}

In Table~\ref{tbl:page1} we show the first page of the JVAS/CLASS results
database.  The entire sample is available as an ASCII text file from 
{\tt http://www.jb.man.ac.uk/research/gravlens}.  This table contains the raw
output from the automapping, and does not include manual remapping of
questionable cases.  Note that the last six
columns contain the modelfitting information for position, flux density,
size and orientation of the Gaussian components.  The axis ratio is often
found to be zero, which indicates a component that it thinks is unresolved
in one dimension.  If the major axis is small (150 mas or less), then this
indicates an unresolved component, while if the major axis is significantly
larger than this, then a zero or very small axis ratio usually indicates a
poorly fitted component.  These pathological cases are most often seen in
cases where a large number of spurious and usually weak components are fit
to uncleaned sidelobes on poorly calibrated data.  For these objects, manual
imaging, self-calibration and more careful modelfitting are being carried out.
The reader should be careful in interpreting the auto-mapping output for 
sources with more than two components for this reason.

Figure~\ref{fig:fluxes} shows a histogram of the 4.85~GHz flux densities of
the complete sample taken from the GB6 survey.  The 30~mJy lower flux 
density limit is clearly visible.  Also shown in Figure~\ref{fig:fluxes} is
the distribution of measured 8.46~GHz flux densities for detections in the
complete sample. 
Sources generally have less detected flux density at 8.46~GHz than the 
4.85~GHz survey flux density. This is for a combination of
reasons but mainly spectral index and the fact that some of the sources
are extended and partially resolved with the VLA in A-configuration.

Histograms of the GB6 vs. NVSS spectral indices and CLASS vs. GB6 are plotted
in Figure~\ref{fig:hista} for those sources in the complete sample that were
detected in CLASS.  The spectral index distribution derived using the observed
8.46~GHz band merged flux densities and the 4.85~GHz GB6 catalogue values is
rather broad, centered around $\alpha \approx -0.6$, with a long tail
steepward.  This behavior might reflect turnover in the spectra, source
variability, resolution effects, a Malmquist-like bias, or likely a
combination of most of the above.  In particular, near the cutoff GB6 flux
density the distribution will likely be dominated by sources scattered upward
in flux density into the sample due to variability or measurement error over
and above those lost downward out of the sample --- this is the Malmquist bias
effect.  In addition, steeper-spectrum sources tend to be
extended and thus are more likely to be resolved in the high-resolution CLASS
observations and to have depressed 8.46~GHz flux densities, further steepening
their apparent integrated spectrum.

\begin{figure*}
\vskip -1cm
\includegraphics[width=16cm]{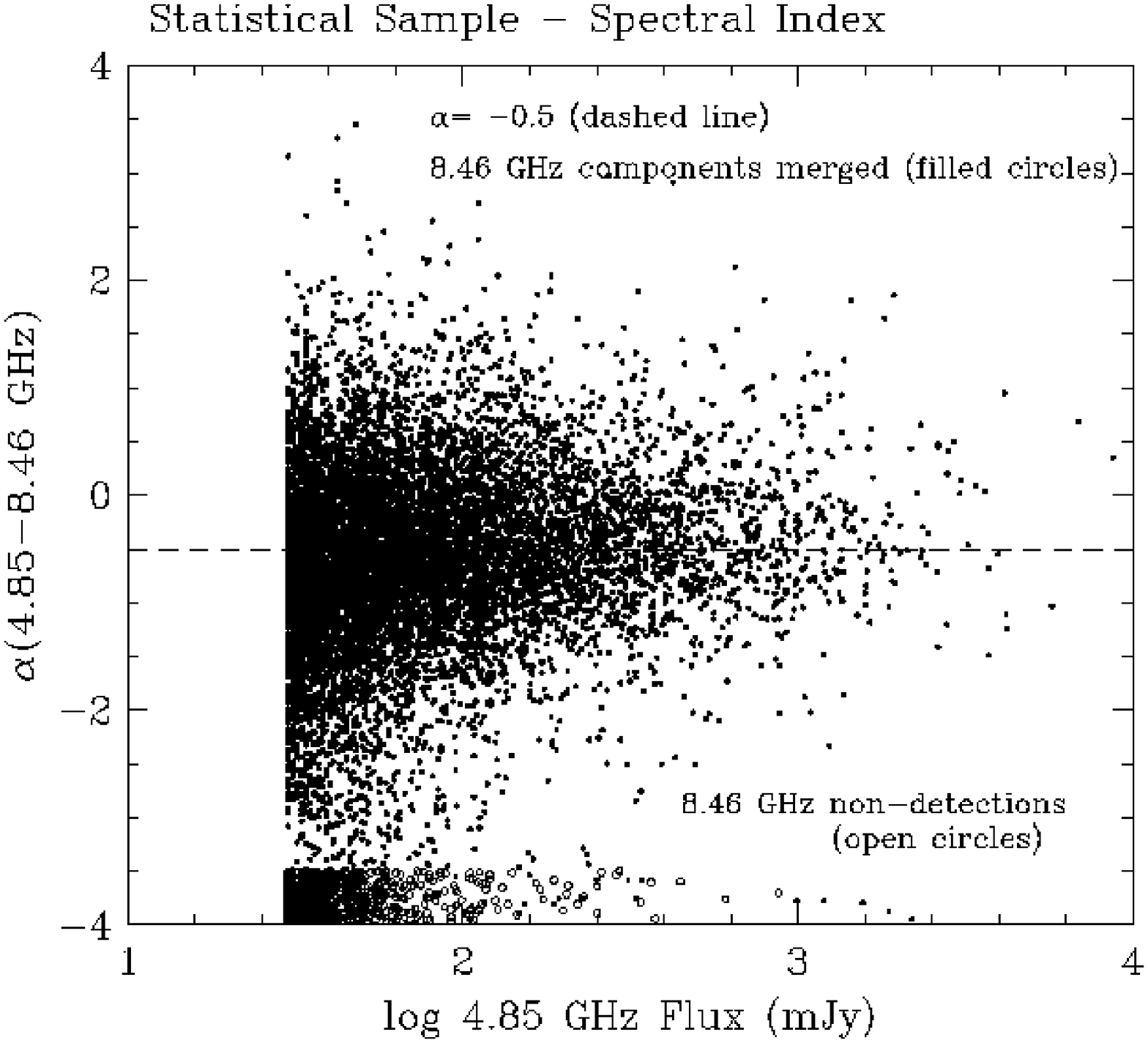}
\vskip -1cm
\caption{Spectral indices between CLASS and GB6 (8.46~GHz and 4.85~GHz) 
plotted versus GB6 4.85~GHz flux density for the complete sample.
As before, the bar of of open circles at the bottom represents sources
in the sample for which no 8.46~GHz detection was made. 
\label{fig:alpha3}}
\end{figure*}

\begin{figure*}
\vskip -1cm
\includegraphics[width=16cm]{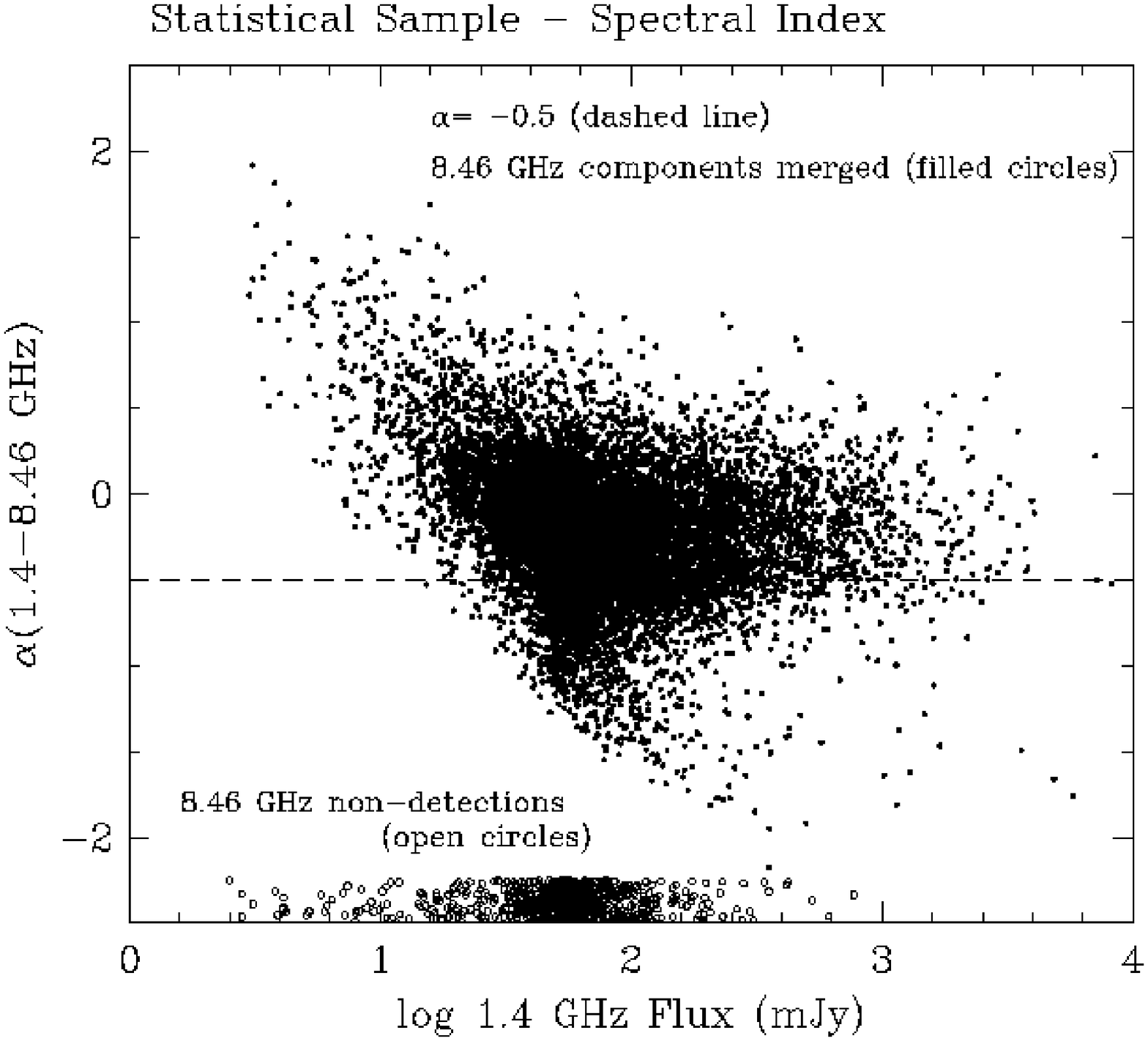}
\vskip -1cm
\caption{Spectral indices between CLASS and NVSS (8.46~GHz and 1.4~GHz) 
plotted versus NVSS 1.4~GHz flux density for the complete sample.
As before, the bar of of open circles at the bottom represents sources
in the sample for which no 8.46~GHz detection was made. 
\label{fig:alpha4}}
\end{figure*}

In Figure~\ref{fig:jclvsgb6} the measured 8.46~GHz flux densities are plotted
versus the GB6 flux density for the complete sample.  The long tail toward
steep spectral indices $\alpha < -0.5$ and non-detections in CLASS are clearly
shown.  A similar plot of CLASS/GB6 (8.46~GHz to 4.85~GHz) versus GB6/NVSS
(4.85~GHz to 1.4~GHz) spectral indices is shown in Figure~\ref{fig:alpha2}.
The spread in CLASS/GB6 spectral index is large.  Furthermore, those sources
with highly inverted ($\alpha > 1.5$) spectra between GB6 and NVSS tend to
have moderate CLASS/GB6 indices.  The CLASS non-detections are again shown as
a bar at the bottom, and these are predominantly sources with GB6/NVSS
spectral index near the cutoff, although there is a significant tail of
sources with inverted low-frequency spectra.  
The CLASS/GB6 spectral index is plotted versus
GB6 flux density in Figure~\ref{fig:alpha3}.  The trend toward larger scatter
in spectral index with lower GB6 flux density is clearly seen.  The upper
envelope with inverted spectra concentrated to lower flux densities at the
lower frequency band is expected, however the excess of steep spectrum sources
at the lower flux densities is worrisome, and likely related to the problem of
non-detections.  Finally, the 8.46~GHz to 1.4~GHz NVSS spectral index is
plotted versus the NVSS flux density in Figure~\ref{fig:alpha4}.  The spectral
index range is tightened up considerably, and the envelopes due to selection
are clearly seen with the non-detections clustering at the limit for the
spectral and flux density cutoffs.  The ``cleaner'' nature of this relation 
is likely due to being well above the NVSS survey flux density limit (reducing
bias), the fact that the CLASS and NVSS surveys are nearly contemporaneous,
and the intrinsic reduction in variability effects at the lower NVSS frequency.

There is a known increase in the rms uncertainty in GB6 flux densities at lower
declination, with the $5\sigma$ limit $S_0$ increasing from less than 20~mJy
at declination $\delta = +20^\circ$ to over 30~mJy at the equator 
(Gregory et al.\ 1996).  To investigate this effect, we plot the
histograms of the CLASS/GB6 spectral index splitting the sample at declination
$+20^\circ$ (Figure~\ref{fig:histb}).  There is a slight widening of the
distribution and shift of the centroid toward steeper spectral index in the
lower declination sub-sample consistent with the effect.  This points toward
problems with GB6 flux densities near the selection cutoff, especially as we
see a significantly increased non-detection rate at lower declinations also
(see below).

\subsection{Detection rate}

The statistics of detections and multiple components found in the
JVAS/CLASS sessions are given in Table~\ref{tbl:stat}.  Of the 16503
targets observed, 14593 were detected in the auto-mapping stage (88.4\%).

In Figure~\ref{fig:nosh} we display the percentage of observed sources which
were not detected in the complete sample, as a function of 5-GHz flux density
and spectral index. It can be clearly seen that the detection rate drops both
with decreasing flux density and steeper spectra, as expected.  In the survey
as a whole, the non-detection rate is much higher (11.6\%) than in the complete
sample (6.7\%).  The extra non-detections arise from a variety of reasons
which have already been described. The main extra contribution is likely to
arise from parts of the survey where the spectral selection was relaxed or
absent (particularly parts of CLASS-1). This is likely to have caused
significantly more contamination by steep-spectrum and resolved sources.  In
addition, a few of the inverted-spectrum non-detections found at lower galactic
latitudes are identified as planetary nebulae.

\begin{figure}
\includegraphics[width=8cm]{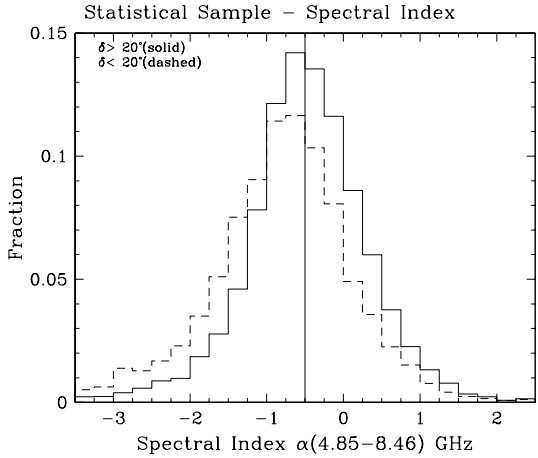}
\caption{Histogram of spectral indices observed between CLASS and GB6 as
a function of declination: Dec $>20^\circ$ (solid) and Dec $<20^\circ$
(dashed).  The vertical axis is the fraction of the total falling in that
bin.  Only sources from the complete sample with CLASS detections are plotted
(with any multiple components merged). 
\label{fig:histb}}
\end{figure}

The 6.7\% non-detection rate in the complete sample is relatively low:
nonetheless a 30-mJy source at 5$\,$GHz should be detectable at $>20\sigma$ at
8.46$\,$GHz, even with a moderately steep radio spectrum.  If we split the
complete sample at declination $+20^\circ$ as we did earlier (see
Figure~\ref{fig:histb}), we find 
that for $\delta > 20^\circ$ the non-detection rate is only 3.8\%, and
rises to 10.4\% for $\delta < 20^\circ$.  Again, this points toward
problems related to the GB6 flux density scale near the cutoff, as
the GB6 rms flux density uncertainty rises sharply at low declinations.  Note
that the non-detection rate for GB6 flux densities above 60~mJy drops
to 2\% overall (46\% of the total sample is above this cutoff).

Of the non-detections, 100 are within the existing regions of the FIRST survey
(Becker, White \& Helfand 1995) and can be investigated in detail.
Just over half of the non-detections (53/100) are due to classical
double radio sources, in which one of the radio lobes is not picked up
within the $70^{\prime\prime}$ search radius in the GB6/NVSS correlation 
(if the second lobe is included, then the source would be dropped from the
sample as the combined spectral index will pass the cutoff).  
Of the remaining 47 sources, 21 are revealed as classical double
sources by FIRST. These are either single NVSS identifications, in the
case of doubles with separation $<45^{\prime\prime}$, or more rarely larger
sources; in either case they appear in the CLASS survey because the
overall spectral index falls just under the $\alpha^{4.85}_{1.4}$ cut.
26 sources remain. Of these, eight are obviously large lobes or extended
structure in which individual hotspots have registered in NVSS,
resulting in inclusion in the CLASS sample.  For eighteen sources there is no
obvious explanation for why they were not detected at 8.46~GHz; they generally
have spectra approaching the spectral index cut, but are unresolved or only
slightly resolved by the FIRST survey, and 6 of these 18 sources have a single
NVSS identification and are unresolved by FIRST.   However, these unexplained
non-detections are most likely to be either very small, relatively
flat-spectrum double sources, highly variable 
flat-spectrum sources, or sources for which the observations failed (although
there is no evidence for the latter). In general, therefore, we can state that
the contamination of the survey by intrinsically steep-spectrum sources
is about 4--5\%, and about 1.5\% of the survey sources are not detected
for reasons that are not clear.

\subsection{Statistics of the northern CLASS-1 region}\label{stats-north}

As mentioned in \S\ref{sample:class1and2}, the Texas 325~MHz survey only
covered the region $\delta < 71\fdg6$ and thus north of this declination there
was effectively no spectral selection in CLASS-1a.  There were 489
87GB-selected sources that were observed in this manner, out of a total of 533
sources in 87GB in this region and flux density of 50~mJy or higher, with the
remaining 44 sources having been previously detected in other surveys.  These
533 sources form a complete sample free of spectral selection that can be used
to check the efficiency of our selection criteria in filtering out
unresolved or complex sources.

Of the 489 CLASS targets in the north region, 350 were detected in the
automatic mapping.  Added to the 44 previous detections in the region,
the total detection rate is 394/533 (73.9\%).  This detection rate is
significantly lower than 
the average of the entire CLASS survey (89\%).  In addition, 155 of the 350
CLASS detected sources were found to have multiple components (44.3\%), a
significantly higher rate than the overall CLASS average (14.3\% of detected
sources in the complete sample).  These statistics are consistent with the
expectations for the spectral selection, which is designed to reduce the
number of extended and intrinsically complex sources in the sample.

\section{Discussion and summary}\label{discuss}

In total, 15.4\% of the entire survey and 13.3\% of the complete sample
(17.4\% and 14.3\% of detected sources respectively) have multiple components
identified in the automapping process.  Some of these multiple sources,
however, will be due to sidelobes and other artifacts of the automapping
process. For further analysis we require an estimate of which sources are
genuine multiple objects.  For lens candidate selection, in particular,
further constraints are applied in total 8.46-GHz flux density, in
primary-to-secondary flux density ratio and in component separation.  In
Figure~\ref{fig:dist} we show the histogram of maximum component separations
in the full sample (the complete sample shows essentially the same
distribution).  The survey is clearly sensitive to multiple components with
separations from 160~mas to 300 arcseconds, and thus our primary goal of
identification of gravitational lens candidates is achievable.  Note that in
Figure~\ref{fig:dist} double-lobe sources, knots within lobes, core-jet
sources, superposition of unrelated sources, artifacts of the automapping
process and real multiple-image lens systems are all mixed together, and
lenses are a tiny fraction of the total number of multiples!  When restricted
to components with flux densities of 2~mJy and above, 15.1\% of all sources
detected and 11.6\% of detected sources in the complete sample were multiple
--- these numbers better represent the fraction of true multiples in the
survey.  Details of candidate selection will be discussed in Paper~2 of this
series.

\begin{figure}
\includegraphics[width=7cm]{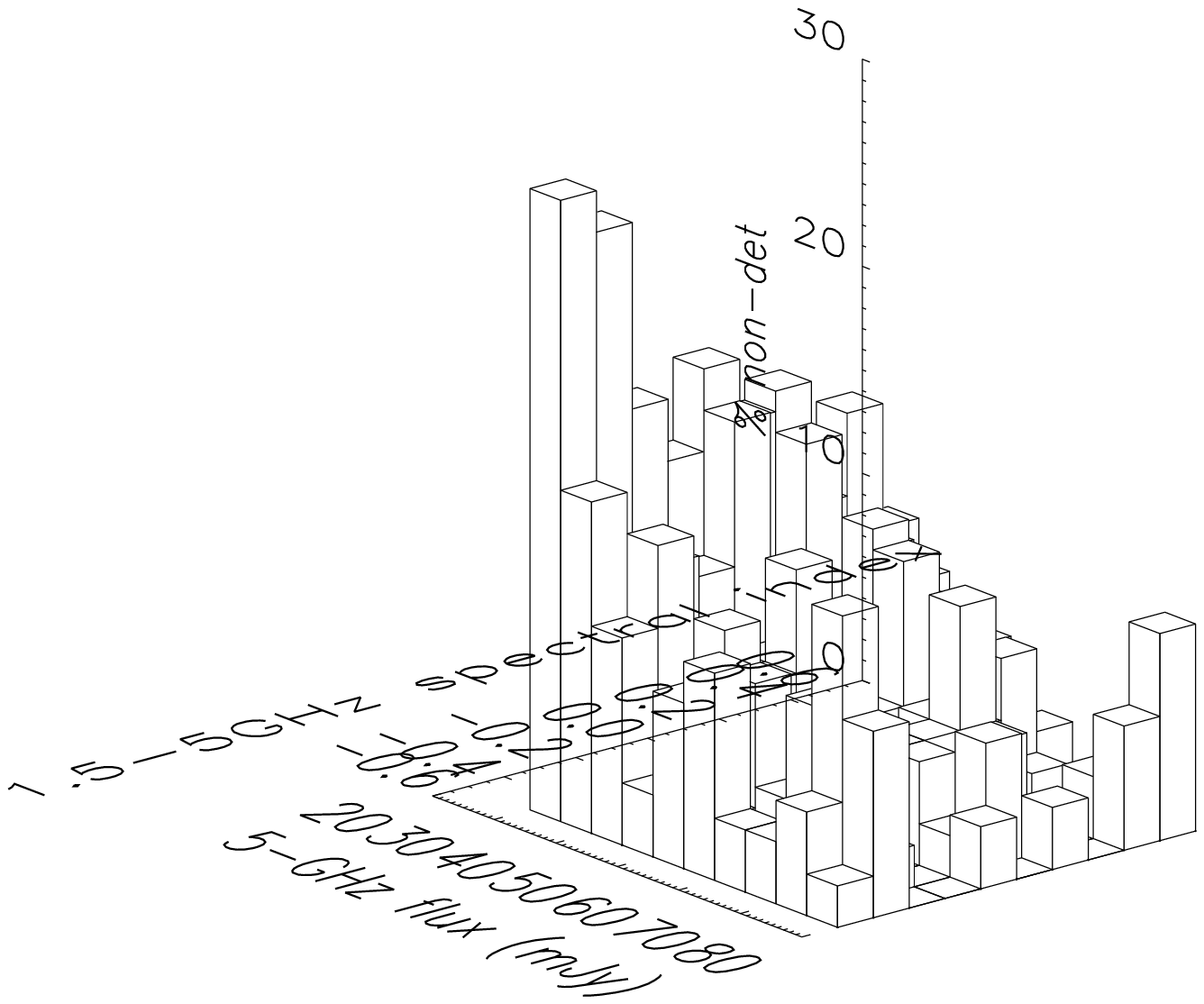}
\vspace{1cm}
\includegraphics[width=7cm]{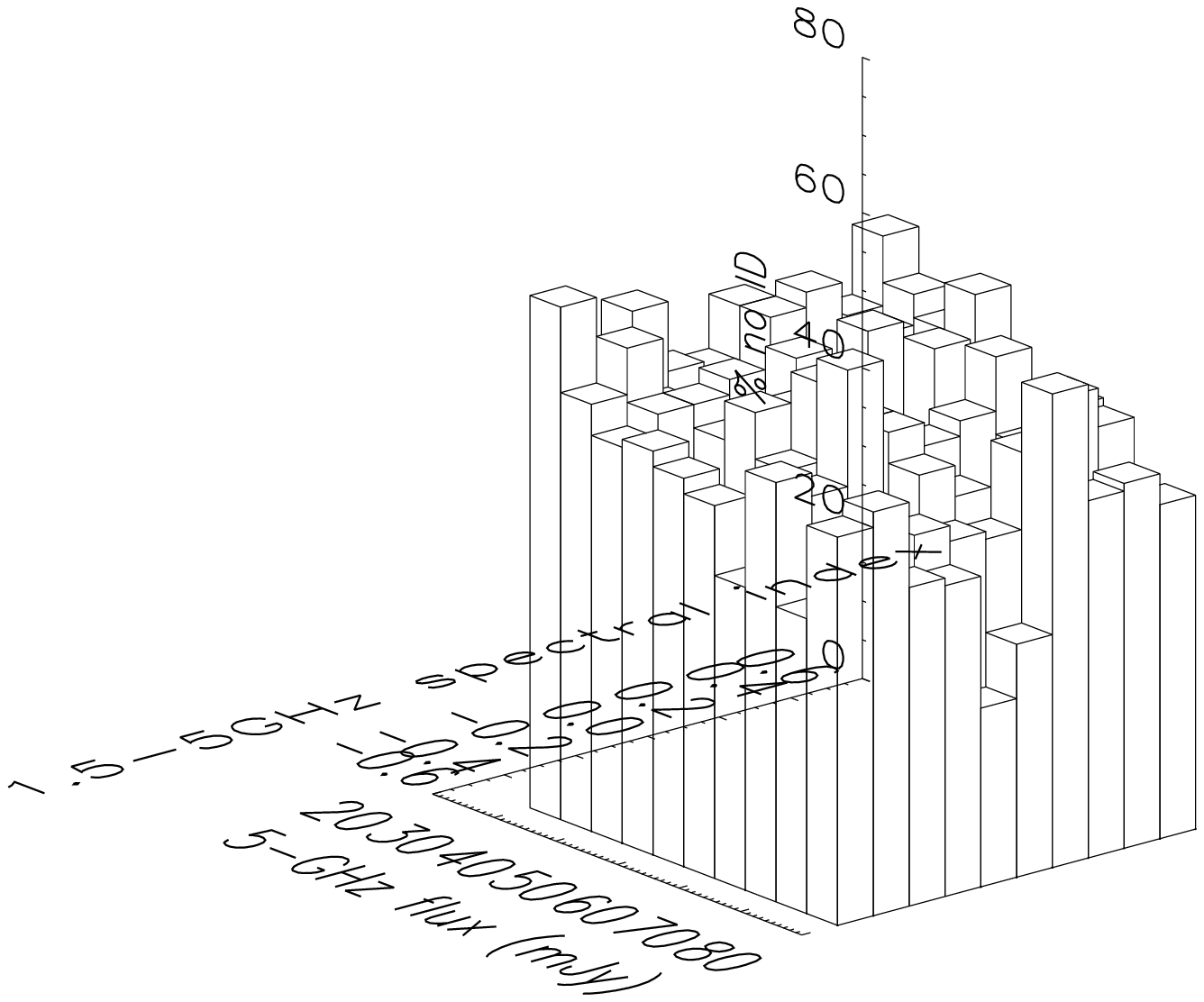}
\caption{Rates of (top) non-detections in 8.46-GHz observations in the full 
JVAS/CLASS sample, and (bottom) sources without optical IDs, in the 
smaller statistically complete sample, plotted as
a function of 1.4-4.85 GHz spectral index and 4.85-GHz flux density. 
Note the strong dependence
of 8.46-GHz detection probability on both radio spectral index and
flux density, whereas the optical detection rate appears to depend mainly
on radio flux density and decreases more gently.
\label{fig:nosh}}
\end{figure}

Optical identifications have been sought from scans of the Palomar Sky Survey
prints carried out by Kibblewhite et al.\ (1984). As expected from earlier
work, the identification fraction of flat-spectrum sources to R$\sim$20
decreases with decreasing flux density, reaching 50\% at $\sim50\,mJy$ and
decreasing further thereafter. Redshift surveys are in progress to identify
sources at these flux density levels (Falco, Kochanek \& Mu\~noz 1998; Marlow
et al.\ 2000).  Figure~\ref{fig:nosh} shows the optical identification rate to
this level; it can be seen that the decrease with 5-GHz flux density is
monotonic, and does not depend strongly on radio spectral index.

In summary, we have presented radio observations of a complete sample of
flat-spectrum radio sources with the VLA. A total of 16503 distinct target
sources have been observed, of which 11685 form a complete sample with the
properties $S_{\rm 5GHz}>30\,$mJy, $|b|>10^{\circ}$,
$\alpha^{4.85}_{1.4}>-0.5$.  Contamination by steep-spectrum sources is around
4\% in the complete part of the survey. Further papers in the series will
discuss the selection of candidate gravitational lenses and astrophysical
inferences from the statistics of this sample.

\section{Acknowledgments}

STM was supported by a R.A. Millikan Fellowship while at Caltech from
1992--1995, and by an Alfred P. Sloan Fellowship at the
University of Pennsylvania.  The CLASS survey at Caltech was supported by NSF
grant AST-9117100, and RDB acknowledges the additional support of NSF
AST-9529170, AST-9900866 and NASA NAG 5-7007.  
This research was supported in part by European
Commission TMR Programme, Research Network Contract ERBFMRXCT96-0034
``CERES''.  Finally, but not least, we thank the staff of the VLA for their
assistance during our observing runs, and for providing the instruments that
made this work possible.


\begin{thebibliography}{}

\bibitem[Augusto et al.(1998)]{au98}
Augusto P., Wilkinson P.N., Browne I.W.A.\ 1998, MNRAS, 299, 1159 

\bibitem[Augusto \& Wilkinson(2001)]{auwi01}
Augusto P. \& Wilkinson P.N.\ 2001, MNRAS, 320, L40

\bibitem[Augusto et al.(2001)]{au01}
Augusto P., et al., 2001, MNRAS, 326, 1007

\bibitem[Baars et al.(1977)]{ba77}
Baars J.W.M., Genzel R., Pauliny Toth I.I.K., Witzel A., 1977,  A\&A, 61, 99

\bibitem[Becker, White \& Helfand(1995)]{be95}
Becker R.H., White R.L., Helfand D.J., 1995, ApJ, 450, 559

\bibitem[Biggs et al.(1999)]{bi99}
Biggs A.D., Browne I.W.A., Helbig P., Koopmans L.V.E., Wilkinson P.N.,
Perley R.A., 1999, MNRAS, 304, 349

\bibitem[Burke(1989)]{bu89} 
Burke B.F., 1989, in "Gravitational Lensing", ed. Y. Mellier 
et al.\ 1989 p.127
 
\bibitem[Browne et al.(1998)]{br98} 
Browne I.W.A., Wilkinson P.N., Patnaik A.R., Wrobel J.M., 
1998, MNRAS, 293, 257

\bibitem[Browne et al.(2002)]{br02}
Browne I.W.A., et al., 2002, MNRAS, submitted (Paper 2)

\bibitem[Chae, Mao \& Augusto(2001)]{ch01}
Chae K.-H., Mao S., Augusto P.\  2001, MNRAS, 326, 1015

\bibitem[Chae et al.(2002)]{ch02}
Chae K.-H, et al., 2002. Phys.\ Rev.\ Lett., submitted (Paper 3)

\bibitem[Cohn et al.(2001)]{co01} 
Cohn J.D., Kochanek C.S., McLeod B.A., Keeton C.R., 2001, ApJ, 554, 1216

\bibitem[Condon et al.(1998)]{co98} 
Condon J.J., Cotton W.D., Greisen E.W., Yin Q.F., Perley R.A., Taylor
G.B., Broderick J.J., 1998, AJ, 115, 1693

\bibitem[Douglas et al.(1980)]{do96}
Douglas J.N., Bash F.N., Bozyan F.A., Torrence G.W., Wolfe C., 1996, AJ, 
111, 1945

\bibitem[Falco, Kochanek \& Mu\~{n}oz(1998)]{fal98} 
Falco E.E., Kochanek C.S., Mu\~{n}oz J.M., 1998, ApJ, 494, 7

 
\bibitem[Fassnacht et al.(1999a)]{fa99a} 
Fassnacht C.D., et al., 1999a, AJ, 117, 658
 
\bibitem[Fassnacht et al.(1999b)]{fa99b} 
Fassnacht C.D., Pearson T.J., Readhead A.C.S., Browne I.W.A.,
Koopmans L.V.E., Myers S.T., Wilkinson P.N., 1999b, ApJ, 527, 498

\bibitem[Fassnacht et al.(2002)]{fa02} 
Fassnacht C.D., Xanthopoulos E., Koopmans L.V.E., Rusin, D., 2002, ApJ,
submitted
 
\bibitem[Gregory \& Condon(1991)]{gr91} 
Gregory P.C. \& Condon J.J. 1991, ApJS, 75, 1011
 
\bibitem[Gregory et al.(1996)]{gr96} 
Gregory P.C., Scott W.K., Douglas K., Condon J.J., 1996, ApJS, 103, 427

\bibitem[Hewitt et al.(1992)]{he92} 
Hewitt J.N., Turner E.L., Lawrence C.R., Schneider D.P., Brody J.P., 1992,
AJ, 104, 968 

\bibitem[Jackson et al.(1995)]{ja95}
Jackson N., et al., 1995, MNRAS, 274, L25

\bibitem[Jackson et al.(1998)]{ja98}
Jackson N., et al., 1998, MNRAS, 296, 483


\bibitem[Keeton \& Madau(2001)]{ke01}
Keeton C.R., Madau P., 2001, ApJL, 549, L25

\bibitem[Kibblewhite et al.(1984)]{ki84} 
Kibblewhite E.J., Bridgeland M.T., Bunclark, 
P.S., Irwin M.J., 1984, Astronomical Microdensitometry Conference,
GSFC, 1983, ed. D.A. Klingles

\bibitem[King et al.(1997)]{ki97}
King L.J., Browne I.W.A., Muxlow T.W.B., Narasimha D.,
Patnaik A.R., Porcas R.W., Wilkinson P.N., 1997, MNRAS, 289, 450

\bibitem[Kochanek(1995)]{ko95}
Kochanek C.S., 1995, ApJ, 445, 559

\bibitem[Kochanek(1996)]{ko96}
Kochanek C.S., 1996, ApJ, 473, 595

\bibitem[Koopmans \& de Bruyn(2000)]{kdb00}
Koopmans L.V.E., de Bruyn A.G., 2000, A\&A, 358, 793

\bibitem[Koopmans \& Fassnacht(1999)]{ko99}
Koopmans L.V.E., Fassnacht C.D., 1999, ApJ, 527, 513
 
\bibitem[Koopmans et al.(1998)]{ko98}
Koopmans L.V.E., et al., 1998, MNRAS, 303, 727

\bibitem[Koopmans et al.(2000a)]{ko00a}
Koopmans L.V.E., de Bruyn A.G., Xanthopoulos E., Fassnacht C.D., 
2000a, A\&A, 356, 391

\bibitem[Koopmans et al.(2000b)]{ko00b}
Koopmans L.V.E., et al., 2000c, A\&A, 361, 815

\bibitem[March\~{a} et al.(2001)]{mm01}
March\~{a} M.J., Caccianiga A., Browne I.W.A., Jackson  N., 2001,
MNRAS, 326, 1455

\bibitem[Marlow et al.(1999)]{ma99} 
Marlow D.R., et al., 1999, AJ, 118, 654

\bibitem[Marlow et al.(2000)]{ma00} 
Marlow D.R., Rusin D., Jackson N., Wilkinson P.N., Browne I.W.A.,
Koopmans L.V.E., 2000a,  AJ, 119, 2629

\bibitem[Marlow et al.(2001)]{ma01} 
Marlow D.R., et al., 2001,  AJ, 121, 619


\bibitem[Myers et al.(1995)]{my95} 
Myers S.T., et al., 1995,  ApJL, 447, L5

\bibitem[Myers et al.(1999)]{my99} 
Myers S.T., et al., 1999, AJ, 117, 2565

\bibitem[Patnaik et al.(1992a)]{pa92a} 
Patnaik A.R., Browne I.W.A., Wilkinson P.N., Wrobel J.M., 1992a, 
MNRAS, 254, 655

\bibitem[Patnaik et al.(1992b)]{pa92b} 
Patnaik A.R., Browne I.W.A., Walsh D., Chaffee F.H., Foltz C.B.,
1992b, MNRAS, 259, 1

\bibitem[Patnaik et al.(1995)]{ppb95}
Patnaik A.R., Porcas R., Browne I.W.A., 1995, MNRAS, 274, L5

\bibitem[Phillips et al.(2000a)]{ph00} 
Phillips P.M., et al., 2000a, MNRAS, 319, L7

\bibitem[Phillips et al.(2001)]{ph01} 
Phillips P.M., et al., 2001, MNRAS, 328, 1001

\bibitem[Rengelink et al.(1997)]{re97}
Rengelink R.B., Tang Y., de Bruyn A.G., Miley G.K., Bremer M.N., 
Roettgering H.J.A., Bremer, M.A.R., 1997, A\&AS, 124, 259
 
\bibitem[Rusin \& Ma(2001)]{ruma01} 
Rusin D., Ma C-P., 2001, ApJL, 549, L33

\bibitem[Rusin \& Tegmark(2001)]{rute01} 
Rusin D., Tegmark M., 2001, ApJ, 553, 709

\bibitem[Rusin et al.(2001a)]{ru01a} 
Rusin D., et al., 2001a, AJ, 122, 591

\bibitem[Rusin et al.(2001b)]{ru01b} 
Rusin D., et al., 2001b, ApJ, 557, 594

\bibitem[Rusin et al.(2002a)]{ru02a} 
Rusin D., Norbury M., Biggs A.D., Marlow D.R., Jackson N.J., Browne I.W.A.,
Wilkinson P.N., Myers S.T., 2002a, MNRAS, 330, 205


\bibitem[Shepherd Pearson \& Taylor(1994)]{sh94} 
Shepherd M.C., Pearson T.J., Taylor G.B., 1994, BAAS, 26, 987

\bibitem[Shepherd(1997)]{sh97} 
Shepherd M.C.\ 1997, in Astronomical Data Analysis Software and
Systems VI, ed.\ G.\ Hunt \& H.E.\ Payne, ASP Conference Series, 125, 77

\bibitem[Snellen et al.(1995)]{sn95}
Snellen I.A.G., de Bruyn A.G., Schilizzi R.T., Miley G.K., Myers S.T., 
1995, ApJL, 447, L9

\bibitem[Snellen et al.(2000)]{sn00}
Snellen I.A.G., Mack K.-H., Tschager W., Schilizzi R., 2000, 
Proceedings of the 5th EVN Symposium. Eds. J.Conway, A.Polatidis, R.Booth.
(Onsala Space Observatory, Chalmers Technical University, Gothenburg, Sweden)

\bibitem[Snellen et al.(2001)]{sn01}
Snellen I.A.G., McMahon R.G.,  Dennett-Thorpe J., Jackson N., Mack K.-H., 
Xanthopoulos E., 2001, MNRAS, 325, 1167

\bibitem[Snellen et al.(2002)]{sn02}
Snellen I.A.G., McMahon R.G., Hook I.M., Browne I.W.A., 2002, 329, 700

\bibitem[Sykes et al.(1998)]{sy98} 
Sykes C.M., et al., 1998,  MNRAS, 301, 310

\bibitem[Turner, Ostriker \& Gott(1984)]{tu84} 
Turner E.L., Ostriker J.P., Gott J.R., 1984, ApJ, 284, 1

\bibitem[White \& Becker(1992)]{wh92} 
White R.L., Becker R.H., 1992, ApJS, 79, 331
 
\bibitem[Wilkinson et al.(1998)]{wi98} 
Wilkinson P.N., Browne I.W.A., Patnaik A.R., Wrobel J.M., Sorathia B., 
1998, MNRAS, 300, 790

\bibitem[Winn et al.(2000)]{wi00} 
Winn J.N., et al., 2000, AJ, 120, 2868

\bibitem[Xanthopoulos et al.(1998)]{xa98} 
Xanthopoulos E., et al., 1998, MNRAS, 300, 649


\end{thebibliography}
\end{document}